\documentclass[a4paper,showpacs]{iopart}
\usepackage{amsfonts}
\usepackage{graphicx}
\usepackage{bm}
\usepackage{soul}
\def\opket#1#2 {  {#1} |{#2}\rangle }

\def\d{{\rm d}}

\def\dydxh#1#2{ \partial{#1} / \partial {#2} }
\def\dydxv#1#2{ {{\partial #1} \over {\partial #2}} }

\def\e{{\rm e}}

\def\eps{\varepsilon}

\def\f0{\bar{f}}

\def\g0{\bar{g}}

\def\h0{\bar{h}}

\def\Im{{\rm Im}}

\def\J{{\cal J}}

\def\K2{{\cal K}}

\def\l{{\ell}}

\def\M{{\cal M}}
\def\max{{\rm max}}

\def\min{{\rm min}}

\def\n{{\hat{\bf n}}}

\def\p{{\bf p}}

\def\rmi{{\rm i}}

\def\Re{{\rm Re}}

\def\S{{\cal{S}}}

\def\S{{\hat{\cal S}}}

\def\VV{{V}}

\def\V0{\bar{\VV}}
\def\WW{{W}}

\def\W0{\bar{\WW}}

\def\x{{\bf x}}

\def\sbar{\big/}
\def\dbar{ {\,\sbar\!\sbar} }
\def\dbarr{ {\,\sbar} }

\def\opsket#1#2#3 {  {#1} \dbar_{#2}{#3}\rangle }

\def\opssket#1#2#3 {  {#1} \dbarr_{#2}{#3}\rangle }

\begin{document}

\title[Quality factors of deformed dielectric
cavities]{Quality factors of deformed dielectric cavities}

\author{Michael M White and Stephen C Creagh}

\address{School of Mathematical Sciences, University of Nottingham,
University Park, Nottingham, NG7 2RD, UK}
\ead{pmxmw@nottingham.ac.uk}

\begin{abstract}
An analysis is provided of the degradation that arises in the
quality factor of a whispering gallery mode when a circular or spherical
dielectric cavity is deformed. The large quality factors  of
such resonators are important to their use in applications such
as sensors, wavelength filters or lasers. Yet a straightforward
analysis of the effect of shape deformation on quality factors
cannot given because the underlying complex ray data demanded
by a standard eikonal approximation frequently does not exist.
In this paper we exploit an approach that has been successfully
used elsewhere to describe the strong directional emission of
such systems, based on a perturbative treatment of the relevant
complex ray families. Applicable when the radial perturbation is
formally of the order of a wavelength, the resulting approximation
successfully describes changes to the quality factor using the
ray geometry in a neighbourhood of a discrete set of escaping rays
guiding the directions of maximum emitted intensity.

\end{abstract}


\section{Introduction}

Optical microcavities are widely exploited in applications such
as lasers, sensors and wavelength filters
\cite{Nockel97,Vahala03,Matsko06,Ilchenko06}. A key underlying feature
in such technologies is that these devices support whispering gallery
modes, which, corresponding to ray families confined by  total internal
reflection to the interior of the resonator, are very long lived.
An important practical consideration is then to be able to
characterise the quality factor of such systems. Geometry is
a particularly important factor in this regard. Deformation from
perfectly circular or spherical geometry, while leading to the often
desirable feature of directional emission, typically degrades the
quality factor itself \cite{Gorodetsky06,Gorodetsky07}. Here we treat
the case of smooth shape
deformations but note that qualitatively similar features may also be seen
in microcavities perturbed by notches \cite{Boriskina06} or 
internal scatterers \cite{Hales11,Dettmann09,Dettman08}.

In this paper we focus particularly on the regime of weak
deformation and offer analytical models for the quality factor
in that limit.
Whether deliberately engineered or arising as a consequence
of manufacturing tolerances, even very weak deformations can be
be seen to have a dramatic effect on the external field of these
systems and as a consequence (we will find) on the quality
factor.
Despite the fact that the internal modes and associated
real ray dynamics are hardly qualitatively changed from the corresponding exact
circular or spherical limit, a straightforward application of eikonal
theory frequently fails for such systems because natural
boundaries \cite{creagh10:757916,Percival81,Percival82}
may prevent one from calculating the complex ray data required
by an approximation of the external, evanescent field. Here we exploit
an approach which has successfully been used elsewhere to characterise the
directional emission of such systems \cite{CreaghWhite12,creaghellipse10}.
Rather than demand exact
solutions of the eikonal equation governing ray dynamics, which
natural boundaries may prevent us from continuing sufficiently
far into the complex domain, we instead construct approximate ray
families, calculated using canonical perturbation theory.

\begin{figure}[h]
\centering
\includegraphics[width=0.5\textwidth]{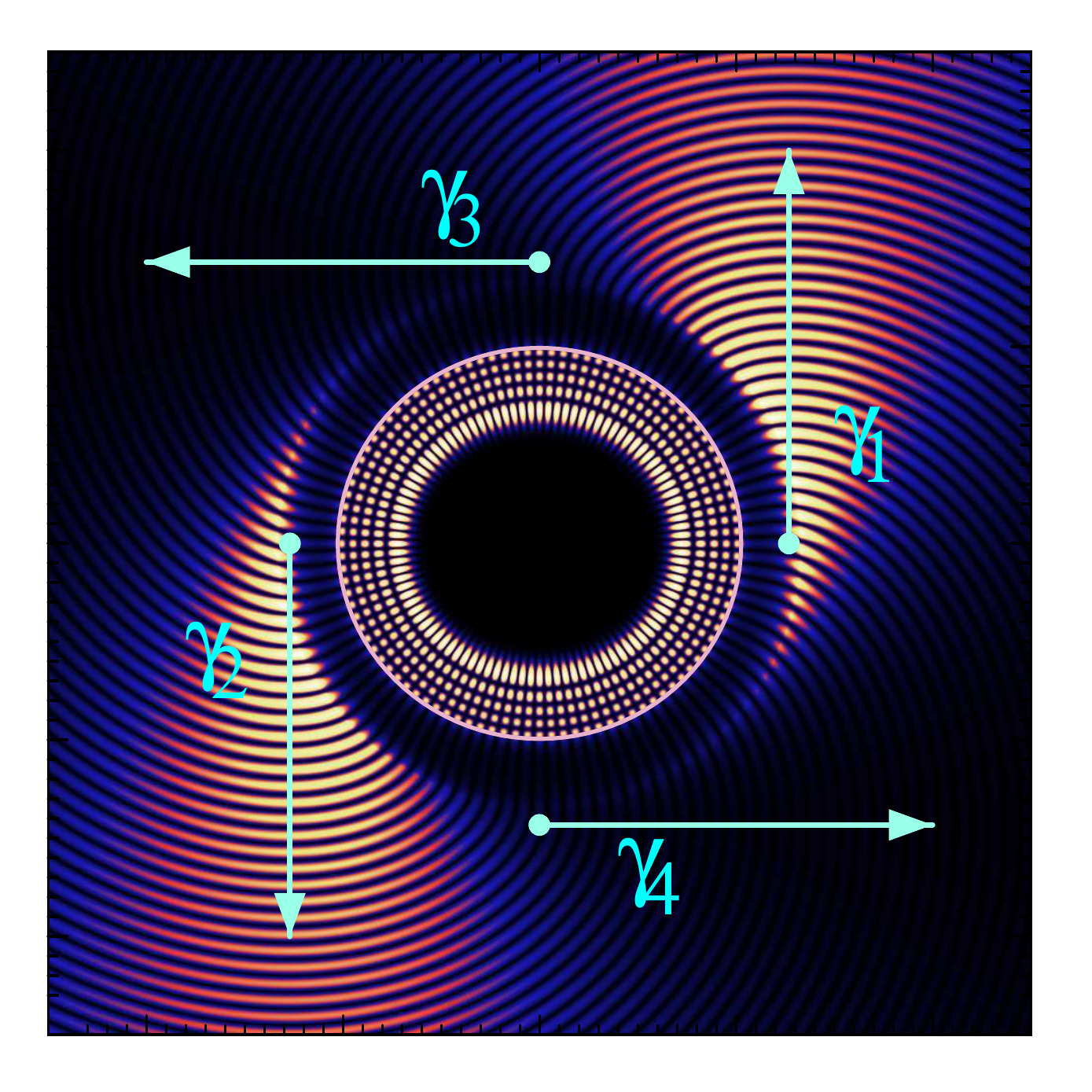}
\caption{The external field is shown for a whispering gallery mode
in an elliptical cavity with refractive index $n=2$ and focal
half-length $c=r_0/4$ where $ab=r_0^2$ is the area.
This mode has azimuthal quantum number $m=50$ and $5$ radial
nodes (with $k_rr_0\approx 37.5$) and we note that the external
field is exaggerated graphically to make it visible.
Superimposed are the real escaping
rays $\gamma_1$ and $\gamma_2$ guiding waves of maximum intensity and the
real rays $\gamma_3$ and $\gamma_4$ guiding waves of minimum intensity.
The geometry of the ray family around these particular escaping rays
determines the $Q$ factor.}
\label{fig:ellipseevan}
\end{figure}

Once the external field has been successfully approximated in
this manner, the quality factor itself may be calculated using Herring's
formula \cite{Wilkinson1986}, which represents it as a flux integral measuring
the overall rate of escape from the resonant mode. A straightforward
evaluation of this flux integral, following a truncation to first order
of a perturbative approximation of ray families, is shown in explicit
numerical examples to describe significant degradations of the quality
factor in problems where the deformation seems at first sight to be
very weak. It should be emphasised that, although the underlying ray
calculation is perturbative, it successfully described emitted waves
that are strongly altered by the deformation and applies when perturbation
of the wave solutions itself is not valid \cite{Dubertrand08}.

When approximated for somewhat larger deformations
using saddle-point approximation, one finds that the dominant
contribution to the Herring integral is associated with a discrete set
of rays escaping to infinity along the directions of greatest emitted
intensity (such as $\gamma_1$ and $\gamma_2$ in Fig.~\ref{fig:ellipseevan}).
Dynamically, the distinguishing feature of these rays is
that they propagate with real direction cosines, whereas one finds
generically in deformed resonators that escaping rays are slightly
complex.
In terms of phase-space geometry, the escaping ray family
locally forms a complex Lagrangian manifold and the dominant
real escaping rays form the intersection of this Lagrangian manifold
with its complex conjugate.
For sufficiently large deformations, the
leading order approximation of the quality factor is determined by
a variant of Wilkinson's formula \cite{Smith06,Wilkinson1986},
which expresses it in terms of
canonically invariant measures if the transversality of this
intersection.

Although a primitive application of the steepest descents
approximation fails in the limit of very small perturbation,
we also offer a uniform analysis allowing one to interpolate
smoothly from the case of small deformation, where simpler
perturbative results apply, to larger deformations where
Wilkinson's formula can be used. This is shown in particular
to give a good description of the quality factor of
elliptical cavities, where the existence of exact invariants
allows us to evaluate the relevant complex ray data for a
wide range of deformations.

We conclude by summarising the content of the paper. We begin in Sec.~2
by setting out the basic problem to be solved and the notation used.
We also describe in general terms how Herring's formula can be used
to calculate the quality factor and outline the main approaches to
its approximation in the cases of weak deformation, where completely
perturbative results are appropriate, and of moderate deformation
where the Herring flux integral may be tackled using approximation
by the method of steepest descents. A more detailed description of
these approaches, and their results, are offered in subsequent
sections. Sec.~3 describes the perturbative approximation which
may be applied to weak but generic deformations. Sec.~4 describes
the asymptotic development of the Herring integral appropriate to
larger deformations and applies this approach in particular to the
special case of elliptic cavities, where the requisite complex ray data may be
explicitly characterised, even for larger deformations. Finally,
conclusions are offered in Sec.~5.

\section{Notation, Herring's formula and its approximation}
\label{sec:setup}

\subsection{Notation}
We consider a two-dimensional scalar model governed by the Helmholtz
equation
\[
-\nabla^2\psi = n^2(\x)k^2\psi.
\]
(We expect that the basic underlying approach offered here, and the
general conclusions reached, will apply also to higher dimensional
problems or vector formulations, but we restrict our detailed
discussion to two-dimensional scalar waves for clarity of exposition.)
Denote the region occupied by the cavity by $\Omega$ and
its boundary by $\partial\Omega$, so that
the refractive index changes discontinuously from a constant value
$n>1$ in $\Omega$ to $n=1$ outside $\Omega$.
The resonances we treat are examples of Gamow-Siegert  states
\cite{Gamow28,Siegert39}, and are solutions
of the Helmholtz equation subject to outwards-radiating boundary conditions
at infinity. Such resonances are found for discrete complex wavenumbers
\[
k = k_r-\frac{\rmi}{2}\kappa
\]
with a negative imaginary part, which is small when the resonances
are long-lived, as is the case for the whispering-gallery modes
we consider. The quality factor
\[
Q = \frac{k_r}{\kappa}
\]
is then large.
The Gamow-Siegert states are not normalisable and instead we adopt
a normalisation convention
\begin{equation}\label{defnorm}
\|\psi\|^2_\Omega\equiv\int_\Omega |\psi(\x)|^2\d\x = 1,
\end{equation}
in which integration is restricted to the cavity's interior.

\subsection{Herring's integral}
Herring's formula is obtained by integrating the identity
\[
\psi^*\nabla^2\psi-\psi\nabla^2\psi^* = 2\rmi\kappa k_r\psi^*\psi
\]
over a region $R$ containing $\Omega$, and using Green's identities to
get
\begin{equation}\label{Herring}
\kappa = \frac{1}{2\rmi k_r\|\psi\|_R^2}\int_\Sigma
\left(\psi^*\dydxv{\psi}{n}-\psi\dydxv{\psi^*}{n}\right)\d s,
\end{equation}
where $s$ represents arc-length along $\Sigma\equiv\partial R$ and
where $\|\psi\|_R^2$ is defined as in (\ref{defnorm}), but with
integration restricted to $R$ instead of $\Omega$.
The result is, of course, independent of the choice of region
$R$, although some restriction on this choice makes the
approximation that follows easier. In particular, we choose
the section $\Sigma$ to
be far enough outside $\Omega$ that the rays underlying
an eikonal approximation of $\psi$ are approximately real (see below),
but close enough to $\Omega$ that the exponentially small contribution to
$\|\psi\|_R^2$ from the exterior of $\Omega$ can be neglected,
allowing us to approximate
\[
\|\psi\|_R^2\approx\|\psi\|_\Omega^2=1.
\]
The Herring integral (\ref{Herring}) then measures the flux
of the quasinormalised state $\psi$ escaping across the section
$\Sigma$. We now outline how an asymptotic approximation of it
is determined in terms of the geometry of the
ray family underlying $\psi$.

\subsection{Approximation of the wavefunction}
In the case of whispering gallery modes in a circular cavity, the
wave field immediately outside the resonator decays exponentially
in the radial direction and is approximated by a family of rays whose
radial momentum is purely imaginary. Far enough outside the resonator,
however, a caustic is encountered beyond which the rays once
again propagate with real radial momentum and the wavefunction is
oscillatory, albeit with an exponentially small amplitude following
the decay that occurs through the evanescent band immediately
outside the resonator.
We consider the case of deformed resonators
sufficiently close to circular that a deformation of this ray family
can be defined which provides an eikonal approximation
\[
\psi(\x) \approx A(\x)\e^{\rmi kS(\x)}
\]
for the field in the exterior region. Although natural
boundaries may prevent us from defining such a ray family exactly,
we have shown elsewhere that approximate families can be constructed
using canonical perturbation theory which successfully describe
the deformed exterior wave field \cite{CreaghWhite12,Creagh07}.
We explain this construction
in more detail later but for now we simply assume that such a
ray family can be defined, with an overall geometry that is a
deformation of the circular case.

We consider in particular the case where $\x$ is far enough
outside the resonator that the rays of the circular limit are
real. In the deformed case, a symmetry breaking is found to
occur so that the corresponding rays become slightly complex.
That is, even for $\x\in\mathbb{R}^2$, the rays propagate with
complex direction cosines. Then the resulting direction dependence
of the imaginary part $\Im(S(\x))$ of the action function can
strongly modulate the wave intensity $|\psi(\x)|^2$ and one finds
that the Herring integral may be dominated by a few directions
of maximal flux density. We now argue that the condition of maximum
intensity/flux corresponds in terms of ray geometry to the
condition that the escaping rays are real, and provide a general
description of the asymptotic evaluation of the flux integral
around these real ray directions.

\subsection{Approximation of the Herring integral}
To describe the asymptotic evaluation of the Herring integral
(\ref{Herring}) more explicitly, let us write the amplitude
of the wave field in the form
\[
A(\x) = t\sqrt{\rho(\x)},
\]
where $t$ is a transmission coefficient coupling the interior
and exterior solutions, defined so that the remaining density term
takes a standard WKB form.
[We point out that a full WKB treatment of dielectric cavities
requires the inclusion of a phase --- related to the Goos-H\"anchen shift
\cite{Goos47,Unterhinninghofen08} ---
accounting for variation of the
reflection phase along the boundary. To simplify the presentation
as much as possible we assume here that any such phases have been
incorporated into the transmission amplitude $t$.]
Then, at leading order (in an expansion in powers of $1/k$),
Herring's integral becomes
\begin{equation}\label{Herringbis}
\kappa
\approx \int_\Sigma \n\cdot\left(\frac{\p+\p^*}{2}\right) \psi^*\psi\; \d s
\approx \int_\Sigma(\Re\,p_n) |t|^2|\rho|\;\e^{-2k_r\Im S}\; \d s,
\end{equation}
where $\n$ is the unit normal to $\Sigma$, $\p(\x)=\nabla S$ and
$p_n=\n\cdot\p$.

Further simplification of this integral can be achieved  by exploiting the
associated dynamical invariants for modes with an integrable underlying
ray dynamics. Any ray-dynamical system enjoys the
invariant $H=(p_x^2+p_y^2)/2$. An additional invariant, which we denote
by $M$, is assumed to exist for the ray families underlying the modes
we treat. This second invariant
is angular momentum for the circular limit and a deformation of it
for perturbed cavities. In the case of elliptical deformations, an explicit
analytical form will be given for $M$ in Sec.~\ref{sec:ellipse} whereas
for the case of generic deformation, an approximate invariant $M$
is defined implicitly by the canonical perturbation approach we employ
in Sec.~\ref{sec:pert}. Let $\chi$ denote an angle coordinate on $\Sigma$
conjugate to $M$. That is, $\chi$ evolves at a constant rate under
the flow generated by using $M$ as a Hamiltonian and we scale it so
that it has period
$2\pi$. Then, it can be shown that
\[
|\rho|\d s = \frac{\J}{2\pi}\frac{\d\chi}{|p_n|},
\]
where
\[
\J = \left|\dydxv{(\varphi_1,\varphi_2)}{(t,\chi)}\right|
= \left|\dydxv{(H,M)}{(J_1,J_2)}\right|
\]
is a Jacobian relating the variables $(t,\chi)$,
respectively conjugate to $(H,M)$, to regular action-angle
variables $(\varphi_1,\varphi_2,J_1,J_2)$ for the ray family
inside the cavity. The important
feature for now is that $\J$ is constant on ray families.
Herring's integral
can now be written
\begin{equation}\label{Herring2}
\kappa
\approx \frac{\J}{2\pi}\int_\Sigma\frac{\Re \,p_n }{|p_n|}\;
|t|^2\;\e^{-2k_r\Im S}\; \d\chi.
\end{equation}
Significant further approximation of this integral is possible
in two particular regimes.

First, in the case of small generic deformations on a scale $\eps=O(1/k)$,
the imaginary part of the action can be computed perturbatively while the
amplitude variation can be neglected and the quantities $p_n$ and $t$
in this integral replaced by the constant values they take for the
circular limit. Then
\begin{equation}\label{weak}
\kappa \approx \kappa_0\M,
\end{equation}
where
\begin{equation}\label{defM}
\M = \frac{1}{2\pi}\oint\e^{-2k_r\Im(S(\chi)-S_0)}\d\chi,
\end{equation}
where $\kappa_0$ and $S_0(\chi)$ respectively denote the
value of $\kappa$ and the action function for the circular limit.
Further development of this approximation will be offered in Sec.~\ref{sec:pert}
where approximations for $\Im (S(\chi))$ are developed by applying canonical
perturbation theory to the ray families. We also note that the integrand
$\e^{-2k_r\Im(S(\chi)-S_0)}$ provides a leading-order measure of the intensity
of the emitted field and has already been calculated in that context
in \cite{CreaghWhite12}. The simple perturbative procedure pursued
in this paper applies in the generic case where the unperturbed
ray family is not too close to resonances of the ray dynamics. When
ray-dynamical resonances are nearby, a somewhat more involved approximation
of the exterior field is required but even in that case we note that
the $Q$-factor degradation can be evaluated once the exterior field
is known using
\[
\M = \int|\psi_\eps|^2\d\chi/\int|\psi_{0}|^2\d\chi
\]
where $\psi_0$ and $\psi_\eps$ respectively denote the circular
limit and the $\eps$-deformed form of the exterior field.
Explicit calculations are confined in this paper to the simpler
case described by (\ref{defM}), however.

For larger deformations, it is natural to approximate the integral using the
method of steepest descents. For generic deformations,
however, the intervention of natural boundaries means that we often
cannot find the
required ray data. An exception is the case of elliptical cavities where
a complete analytical description of the external ray family can be
achieved. The resulting approximation for this case is described in
detail in Sec.~\ref{sec:ellipse} but here we summarise the main
qualitative features, which also apply to the large-$\eps k$
asymptotics of (\ref{defM}).
Stationary points of the exponent $-2\Im(S) =\rmi (S-S^*)$ in
Herring's integral are obtained for real $\x$ when
\begin{equation}\label{ppstar}
\p = \p^*,
\end{equation}
that is, when the associated escaping rays are \textit{real}.
Generic escaping rays are complex for deformed resonators
(albeit only slightly complex when the resonator is
weakly eccentric). A discrete subset of real escaping rays
can typically be found, however, propagating along the directions
of greatest and least emitted intensity. A steepest-descents
approximation of Herring's integral therefore characterises
the decay rate in terms of the behaviour of the mode around the
directions of greatest concentration of escaping flux, as one
might expect intuitively. Further details of this approach
are developed in Sec.~\ref{sec:ellipse}.

\section{Approximation of Herring's integral for generically perturbed cavities}
\label{sec:pert}

In this section we describe how the weak-deformation approximation
(\ref{weak}) of the Herring integral may be
exploited to describe the degradation of the quality factor
for boundary shapes of the form
\[
r(\theta) = r_0 + \eps r_1(\theta).
\]
Recall that natural boundaries prevent
a systematic calculation of escaping rays in generically deformed
cavities but that, for small enough deformations, approximate
ray families calculated using canonical perturbation theory can
successfully describe the emitted field. We now summarise the main
relevant results from that calculation and use them to evaluate the
integral $\M$ in (\ref{weak}).

\subsection{Calculation of external field using
perturbative approximation of rays}
The first step in finding the complex action function
$S(\x)$ of escaping rays is to calculate the corresponding
action function on the boundary $\partial\Omega$. This must be
done for complex coordinates to allow a continuation to the
farfield by tracing the associated complex escaping rays.

The action is found on $\partial\Omega$ by applying canonical
perturbation theory to the internal dynamics, following the
approach developed by Prange and Zaitsev in the context of
billiard problems \cite{Prange2001,Prange1999}.  Dynamics on the boundary is characterised
by using a perturbative expansion of the chord function
\begin{equation}
L(\theta,\theta')=L_0(\theta,\theta')+\epsilon L_1(\theta,\theta')+\cdots,
\end{equation}
expressing, as a series in the perturbation
parameter $\eps$, the length of the chord connecting points on the boundary
with polar angles $\theta'$ and $\theta$. The leading term
\[
L_0(\theta_,\theta')=2r_0\sin\left|\frac{\theta-\theta'}{2}\right|
\]
is the chord function for a circle of radius $r_0$ and the
first-order term is
\[
L_1(\theta,\theta')=
\sin\left|\frac{\theta-\theta'}{2}\right|
\left(r_1(\theta)+r_1(\theta')\right).
\]
The chord function $L(\theta,\theta')$ serves as a type-one
generating function for the boundary map, expressed in terms of the
polar angle $\theta$ and its conjugate momentum variable $J=p_\theta$.
Note that $(\theta,J)$ serve as action-angle variables for the
unperturbed boundary dynamics. The action-angle
variables $(\bar{\theta},\bar{J})$ for the perturbed system
can then be described by the type-two generating function
\begin{equation}
\S(\theta,\bar{J})=\theta \bar{J}+\epsilon g(\theta,\bar{J})+\cdots,
\end{equation}
where $g(\theta,\bar{J})$ is shown to satisfy the difference equation
 \begin{equation}\label{diff}
 g(\theta+\omega,\bar{J})-g(\theta,\bar{J})
=R(\theta,\bar{J}),
 \end{equation}
where
\[
R(\theta,\bar{J})\equiv L_1(\theta+\omega,\theta)-\langle L_1\rangle
\]
denotes the oscillating part, with respect to $\theta$,
of $L_1(\theta+\omega,\theta)$, with $r_0\cos(\omega/2)=\bar{J}$.
Note that $\bar{J}$
takes a fixed value determined by the quantisation
condition $nk\bar{J}=m$, where $m$ is the azimuthal quantum
number of the unperturbed mode.

This difference equation for $g(\theta,\bar{J})$ is straightforwardly
solved, as a Fourier series for example, for any analytic function
$r_1(\theta)$. Note that the rays escaping to infinity
typically start with complex values of $\theta$. Here we limit,
our perturbative expansion to a first-order truncation, in
which case it suffices to substitute in $g(\theta,\bar{J})$ the
initial conditions for the unperturbed escaping rays of the
circle. For a counterclockwise-rotating mode, these can be
shown to originate at $\theta_0=\chi-\beta=\chi-\pi/2+\rmi\Theta$, where
$\chi$ is the polar angle of a ray in the farfield and
$\beta=\pi/2-\rmi\Theta$ is the angle of refraction
with which it leaves the cavity's boundary.
From Snell's law, $\cosh\Theta=\sin\beta=n\sin\alpha=nJ/r_0$, where
$\alpha$ is the (constant) angle of reflection of the unperturbed internal
rays. If the terms in the Fourier series for $r_1(\theta)$
decay slowly enough, the solution $g(\theta,\bar{J})$ of this Fourier
series may have natural boundaries which lie below the initial
conditions on $\Im(\theta)=\Theta$ for escaping rays. In this
case it is not known how to approximate external field
arbitrarily far outside the resonator. We therefore
restrict our attention to the case where $g(\theta,\bar{J})$ can
be successfully extended in the complex plane as far as $\Im(\theta)=\Theta$.
This is true in particular when $r_1(\theta)$ is a trigonometric
polynomial function of $\theta$, which is the case for the models
used in our numerical illustrations.

Next, the generating function $\S(\theta,\bar{J})$ can also be
shown to determine, at first order, the phase of the wave field
on the boundary. That is, on the boundary,
$\psi
=B(\theta)\e^{n\rmi k(\theta \bar{J} + \eps g(\theta,\bar{J}) + \cdots)}$.
The action along a ray emerging from $\theta$ with polar angle of
propagation $\chi$ is then
\[
S(\x,\bar{J}) = \l(\x,\theta)+n\S(\theta,\bar{J}) +O(\eps^2)
\]
where
\[
\l(\x,\theta)=\l_0(\x,\theta)+\eps\l_1(\x,\theta)\cdots
\]
denotes the length of the ray connecting the point with
polar angle $\theta$ on $\partial\Omega$ to the point $\x$ on $\Sigma$
and where, if required, $\theta$ can be expressed in terms of
$\x$ and $\bar{J}$.

Finally, as the cavity is deformed, the initial
coordinate $\theta$ of the ray to a fixed exterior point changes.
However, the functions $\l(\x,\theta)$ and
$\S(\theta,\bar{J})$ can be shown respectively to satisfy the
generating-function conditions
$\dydxh{\S(\theta,\bar{J})}{\theta} = r_0\sin\alpha$ and
$\dydxh{\l(\x,\theta)}{\theta}=-r_0\sin\beta$. By Snell's law,
the effect on $S(\x,\bar{J})$ of changing $\theta$ therefore
cancels at first order in $\eps$ and we may approximate
\[
S(\x,\bar{J}) = S_0(\x,\bar{J})
+ \eps S_1(\x,\bar{J})+O(\eps^2),
\]
where
\begin{equation}\label{defS1}
S_1(\x,\bar{J}) = \l_1(\x,\theta_0)+ng(\theta_0,\bar{J})
\end{equation}
and $S_0(\x,\bar{J})$ and $\theta_0$ are, respectively, the action function
and the launching angle for the unperturbed limit.

\subsection{Calculating the $Q$-factor}

Noting that rays in the farfield of the unperturbed problem
satisfy $\theta_0=\chi-\pi/2+\rmi\Theta$,
 the $Q$-factor degradation (\ref{defM}) can then be approximated
\[
\M = \frac{1}{2\pi}\oint\e^{-2\eps k_r \Im\, f(\chi)} \d\chi.
\]
where,
\[
f(\chi) = S_1(\x(\chi),\bar{J})
\]
is obtained by evaluating (\ref{defS1}) at a position $\x(\chi)$ on
$\Sigma$ defined by the polar angle $\chi$ and using the unperturbed
launch angle $\theta_0$.

\begin{figure}[h]
\begin{center}
\includegraphics[width=0.75\textwidth]{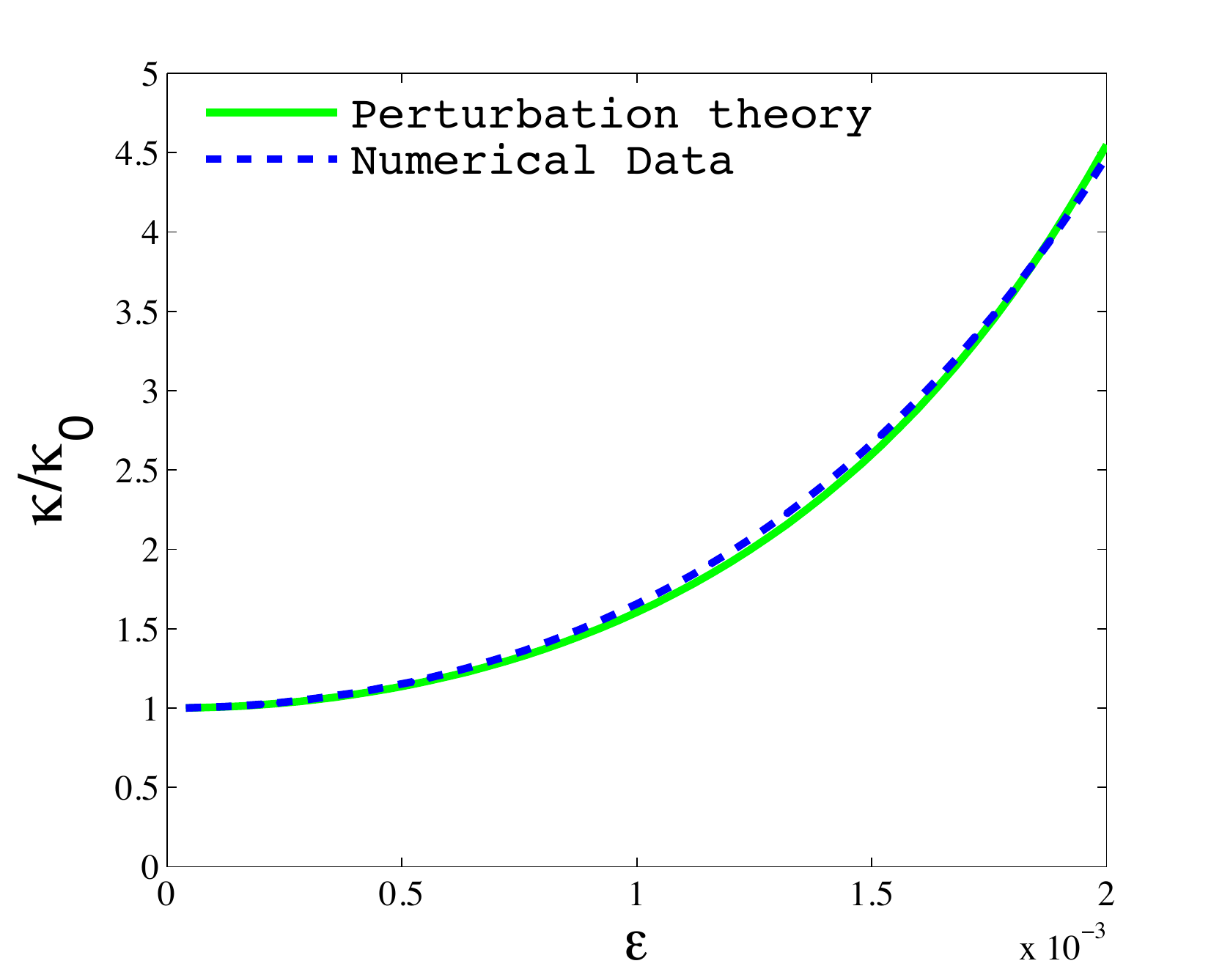}
\end{center}
\caption{The perturbative approximation of $\kappa$ in (\ref{genresult}) is
compared with a numerical evaluation for a deformation of the
form $r=r_0(1+\eps\cos 3\theta)$ of the circle. The example shown
here is a mode with azimuthal quantum number $m=60$ and $6$
radial nodes (with $k_r r_0\approx 45.3$) in a cavity with refractive
index $n=2$.}
\label{fig:kappagen}
\end{figure}

To illustrate the calculation more explicitly we now consider
deformations of the form
\[
r_1(\theta) = r_0\cos N\theta,
\]
where $N$ is an integer. The difference equation (\ref{diff})
has the solution
\[
g(\theta,\bar{J})= \frac{r_0\sin(\omega/2)\sin N\theta}{\tan(N\omega/2)}
\]
in this case, while
\[
\l_1(\x,\theta) = -r_1(\theta)\cos(\chi-\theta),
\]
and it can be shown as a result that
\[
\Im\,f(\chi) = f_0\cos N\left(\chi-\frac{\pi}{2}\right)
\]
where
\[
f_0=
\frac{n r_0\sinh N\Theta\sin(\omega/2)}{\tan(N\omega/2)}-r_0\cosh N\Theta\sinh\Theta
\]
Then,
\begin{equation}\label{genresult}
\M = I_0(2k\eps f_0),
\end{equation}
where $I_0(z)$ denotes a modified Bessel function of the first kind.

A typical implementation of this result is illustrated in
Fig.~\ref{fig:kappagen}. It should be emphasised that even though the
perturbations involved here are quite small (typically of the
order of one part in a thousand in a cavity that is about
30 wavelengths in diameter), the deformation is enough to
significantly change the $Q$-factor --- by a factor of
$4$ or more in the illustration. We also emphasise that, despite
the apparent smallness of the perturbation, natural
boundaries typically prevent us using a straightforward exact
determination of the rays in this example.

\section{Wilkinson's formula and the ellipse}
\label{sec:ellipse}

For larger deformations, where the relevant eikonal phase is a
rapidly-varying function of its arguments, Herring's integral
(\ref{Herring}) invites approximation by the method of steepest descents.
This approach is attractive because it provides a direct
interpretation of the $Q$-factor degradation in terms of simple
geometrical properties of particular escaping rays ---
we will see in fact that
it takes the form of a formula developed by Wilkinson in
the context of tunnel splittings of quantum-mechanical
energy levels \cite{Wilkinson1986}.

For generic deformations such as treated in the previous section,
Wilkinson's formula provides us with a simple ray-geometrical
interpretation of the large-$\eps k$ asymptotics of approximations
such as (\ref{genresult}). It should be acknowledged, however,
that the method will in that case be restricted to moderately large
values of $\eps k$ where the perturbative approach remains valid.
The asymptotic approach  then has the benefit primarily of providing
a useful ray-geometric interpretation of existing calculations rather
than pushing them into fundamentally different regimes.

The case of elliptical deformation, however, is a particularly important
special case where significantly greater deformations  \textit{are}
treatable and where the asymptotic approach has the added value of
extending the results of the previous section to different regimes.

We begin by describing the appropriate generalisation of Wilkinson's
formula to the calculation of $Q$-factor degradation in a more general
context where the escaping rays (or some approximation of them) are
assumed to be known but without making further assumptions. We then apply
this result to the specific case of elliptical cavities where the
known invariants of the underlying ray families allow us to provide
an explicit evaluation of the general result. Finally, we describe at
the end of this section how a uniform evaluation of Herring's
integral allows us to interpolate smoothly between the generic
perturbative results of the previous section and the large-deformation
limit described by Wilkinson's formula.

\subsection{Wilkinson's formula in general}

Wilkinson's formula is obtained by carrying out the steepest-descents
approximation of Herring's integral outlined in
section~\ref{sec:pert}. Recall that we assume the existence (exactly, as in
the ellipse, or more generally following a perturbative approximation)
of an invariant $M$ of the ray family in addition to the usual
ray invariant $H=(p_x^2+p_y^2)/2$ and that the steepest-descents
condition (\ref{ppstar}) is satisfied by a discrete set of real escaping
rays guiding the maximum-intensity waves to infinity.

Then expansion of (\ref{Herringbis}) (or, equivalently, (\ref{Herring2}))
about each of these stationary points
leads to the approximation
\begin{equation}
\label{eq:kappa}
\kappa = \sum_\gamma (\Re\,p_n)|t|^2|\rho|
\left|\frac{k_r}{2\pi\rmi}\frac{\partial^2(S-S^*)}{\partial s \partial s}\right|^{-1/2}
\rme^{-2k_r\Im S},
\end{equation}
where $\gamma$ labels the real rays escaping to infinity along
directions of maximal intensity. Further manipulation of the
amplitude following techniques given in \cite{Littlejohn90} enables it to be expressed in the following
canonically invariant form:
\begin{eqnarray}
\label{eq:kappawilk}
\kappa=\sum_\gamma\frac{|t|^2\J }{ (2\pi)^{3/2}k_r^{1/2}}  \frac{\rme^{-2k_r{\Im}S}}{\sqrt{ |\{M,M^*\}|}}.
\end{eqnarray}
The term in the square root here denotes a Poisson bracket
between the invariant $M$ and its complex conjugate $M^*$.
This Poisson bracket necessarily vanishes on real ray
families, for which $M=M^*$, so the denominator in Wilkinson's
formula effectively provides a (canonically invariant) measure
of the rate at which rays become complex as one moves away from the
particular real rays labelled by $\gamma$. Alternatively it measures
the  transversality of the ray family to its complex-conjugate partner
around their real intersection along $\gamma$. We also note that,
using the fact that $M$ and $M^*$ each Poisson commute with
$H=(p_x^2+p_y^2)/2$, the
Jacobi identity yields $\{\{M,M^*\},H\}=-\{\{M^*,H\},M\}-\{\{H,M\},M^*\}=0$.
Because the flow of $H$ is along rays, this means that $\{M.M^*\}$
is invariant along each real ray $\gamma$ and it therefore doesn't matter
where on $\gamma$ we calculate it.

\subsection{Wilkinson's formula for the ellipse}

As described in \cite{creaghellipse10}, the external field
can be described in detail in the special case of elliptical deformations
by exploiting the existence of explicit analytical
expressions for the invariant $M$ for that case. We now use these results
to offer a detailed evaluation of Wilkinson's
formula for the dielectric ellipse.

We begin by summarising the main geometric characteristics of the
rays of the dielectric ellipse, with a more detailed description
being available in \cite{creaghellipse10}. We denote by $a$ and
$b$ the major and minor semiaxes and by $c$ satisfying $a^2+b^2=c^2$
the half-distance between foci.
Ray families in the
ellipse's interior have, as an invariant,
\[
A(\x,\p)=(xp_y-yp_x)^2-c^2p_y^2,
\]
which expresses in cartesian coordinates the product of angular
momenta about the two foci. The invariant in the exterior takes a
different functional form but can be expressed in terms of the
function $A(\x,\p)$ by matching interior and exterior families
using Snell's law. Our convention is that
the exterior invariant $M$ thus obtained is scaled so that
\begin{equation}
n^2M=A-(n^2-1)c^2\sin^2u.
\end{equation}
Here, $u$ denotes the boundary coordinate from which an
external ray is launched, where in general $u$ parameterises the ellipse
boundary according to $(x,y)=(a\cos u,b\sin u)$.
Because rays guiding the external wave to infinity have
complex starting points on the boundary, $u$ and therefore $M$,
are complex functions of the external phase space coordinates
$(\x,\p)$.
In fact, more detailed calculations \cite{creaghellipse10} show that
\[
\sin u=\frac{-bp_xL+\rmi ap_yQ}{b^2+c^2p_y^2},
\]
where $L=xp_y-yp_x$ denotes angular momentum about the centre of the ellipse,
and $Q\equiv(A-b^2(p_x^2+p_y^2))^{1/2}$.
We can then separate the real and imaginary parts of the
invariant $M$ according to
\begin{equation}\label{ReImM}
n^2M=A-B^2+C^2-2\rmi BC,
\end{equation}
where
\[
B=-c\sqrt{n^2-1}\,
\frac{bp_xL}{b^2+c^2p_y^2}
\quad\mbox{and}\quad
C=c\sqrt{n^2-1}\,\frac{ap_yQ}{b^2+c^2p_y^2}.
\]
In order to implement (\ref{eq:kappawilk}) we must
now find the real escaping rays $\gamma$,  along with their complex
starting points on the boundary, and then evaluate their actions and
amplitudes in (\ref{eq:kappawilk}).

Assuming foci on the $x$-axis, we find that real rays guiding waves of
maximum intensity emerge vertically and have starting points
with $u$-coordinates
\[
u_1=\rmi U_1\quad\mbox{and}\quad
u_2=\pi+\rmi U_1,\quad\mbox{where}\quad
\sinh U_1=\sqrt{\frac{n^2M-b^2}{b^2+n^2c^2}}.
\]
The full coordinates of these starting points are
$(x_0,y_0)=(\pm a\cosh U_1,\pm\rmi b\sinh U_1)$ and the corresponding
rays emerge ($\gamma_1$ and $\gamma_2$ in Fig.~\ref{fig:ellipseevan})
with momentum $\p=(0,\pm 1)$, so that a general point on them has coordinates
$(x,y)=(\pm a\cosh U_1,y_0+ t)$, $t\in\mathbb{C}$.
The imaginary part of the action is
therefore the imaginary part of the $y$ displacement needed
to get to real coordinate space, which is
\[
\Im S = p_y\Im(y_0)=b\sinh U_1.
\]
It remains to evaluate the Poisson bracket in the denominator of
(\ref{eq:kappawilk}), which, from (\ref{ReImM}), takes
the form.
\[
n^4\{M,M^*\}=4\rmi\{A-B^2+C^2,BC\}.
\]
Evaluated on the dominant, vertically escaping rays, this can be
shown after further manipulation
to take the value
\begin{equation}\label{brightPB}
\rmi\{M,M^*\}_{u_1,u_2}=\frac{8 (n^2-1)c^2b}{n^2a^2}\left(M+c^2\right)\sqrt{\frac{n^2M-b^2}{b^2+n^2c^2}}.
\end{equation}
The final ingredient needed to evaluate (\ref{eq:kappawilk})  is
the transmission coefficient, which, following the discussion in
\cite{creaghellipse10}, can be shown to take the form
\begin{equation*}
t=\frac{2\sqrt{n\sin\alpha\cos\beta}}{\cos\alpha+\cos\beta}\e^{\rmi\sigma},
\end{equation*}
where $\alpha$ and $\beta$ are respectively the angles of incidence
and reflection at the boundary point from which the escaping ray
is launched. Our convention in this paper is that this transmission
coefficient also incorporates a phase correction $\sigma$, described
in detail in \cite{creaghellipse10},  that accounts for the variable
reflection phases on the boundary and is related to the
Goos-H\"anchen shift \cite{Goos47,Unterhinninghofen08}.

\begin{figure}[h!]
  \centering
    \includegraphics[width=0.75\textwidth]{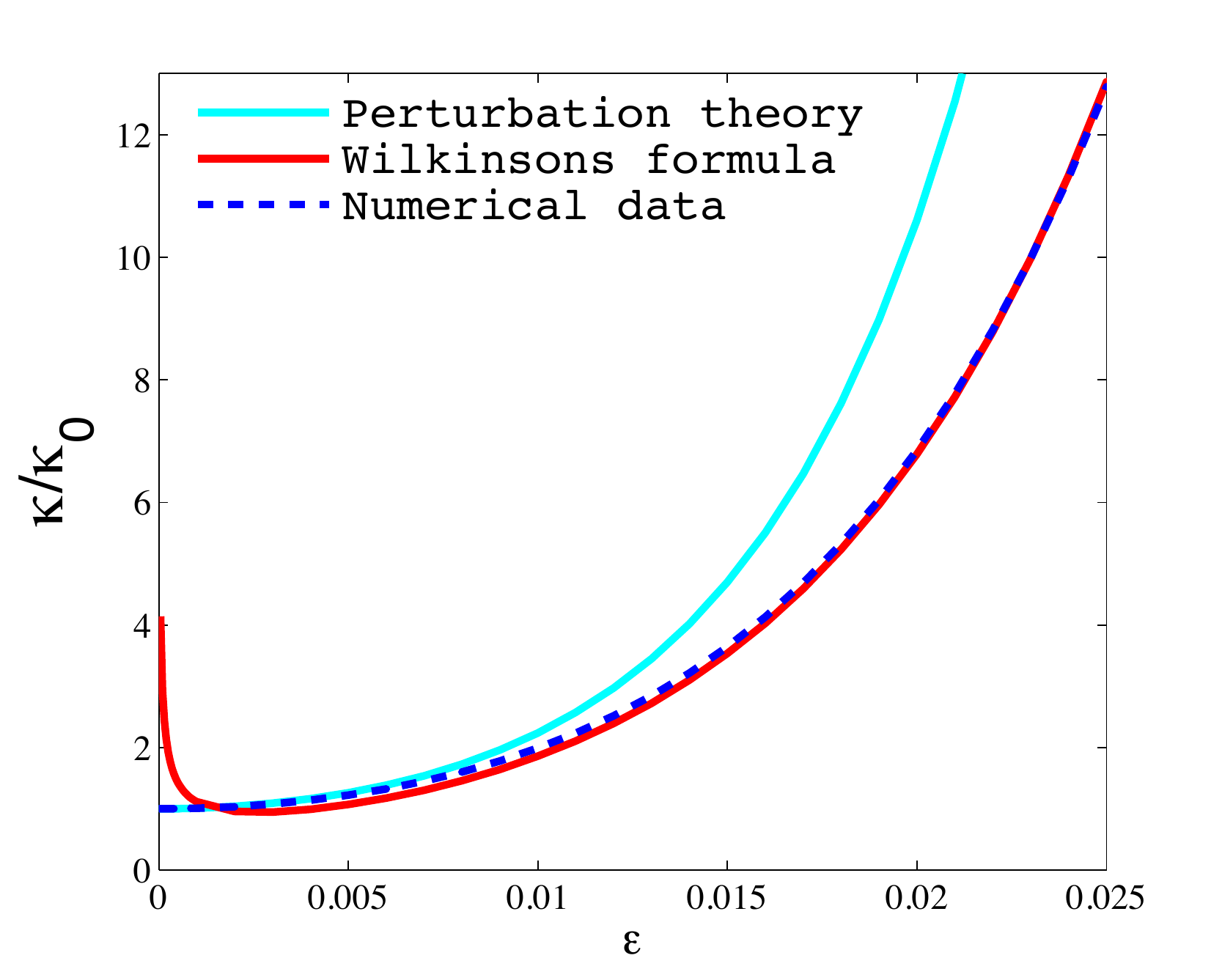}
    \caption{A comparison is given between calculations of $\kappa$
using  Wilkinson's formula (\ref{eq:kappawilk}), the perturbative
approach of the previous section,
and a numerical evaluation. The mode is the same as used in
Fig.~\ref{fig:ellipseevan} and the perturbation parameter $\eps$
is defined so that $b = r_0(1-\eps)$
(so that $c^2=a^2-b^2\approx4\eps r_0^2$).}
    \label{fig:kappawilk}
\end{figure}

A numerical illustration of this result is given in
Fig.~\ref{fig:kappawilk}.
Also shown in the same figure is the perturbative prediction developed
in the previous section. Approximation (\ref{eq:kappawilk}) is seen to
give a good description of the quality factor degradation, and it
works well for deformations strong enough that the perturbative
result (\ref{genresult}) has failed.  Approximation (\ref{eq:kappawilk})
fails, however, for very small perturbations where, in the
integrable circular limit in which all escaping rays become real,
the Poisson bracket $\{M,M^*\}$ vanishes and the amplitude
of (\ref{eq:kappawilk}) diverges. This is because in the regime
of small $k\eps$ the steepest descents approximation used to
approximate $\kappa$ is no longer justified. Here the perturbative
approach of the previous section must be used instead. In the next
subsection we describe how the two limits may be married in a
single, uniform approximation that is valid in both limits.

\subsection{Uniform approximation}
\label{sec:uniform}

A uniform approximation of the $Q$-factor degradation that interpolates
smoothly between the perturbative approximation of Sec.~\ref{sec:pert}
and Wilkinson's formula in (\ref{eq:kappawilk}) is now derived.
The method here follows a uniform treatment in \cite{Ullmo95,Sieber1997}
of the contribution of bifurcating periodic orbits to the trace formula.

We return to the version of Herring's integral given in
(\ref{Herring2}) and seek a coordinate transformation
$s\to\chi$ which is so that the imaginary action takes the form
\[
\Im S = \bar{K}+\Delta K \cos2\chi,
\]
where $\bar{K}$ and $\Delta K$ are constants. The form chosen here
exploits the symmetry of the problem with respect to
$u\to u+\pi$ and the variable $\chi$ thus defined coincides at
leading order with the polar angle used in the perturbative Herring
integral (\ref{weak}), but deviates from it at higher deformation.

The constants $\bar{K}$ and $\Delta K$ are determined by
evaluating the minimum and maximum values $K_\min=\bar{K}-\Delta K$
and $K_\max=\bar{K}+\Delta K$ of $\Im\,S$.  We have already
established that the minimum value $K_\min=b\sinh U_1$ is
achieved by the real rays escaping along the directions of greatest
intensity. The maximum value $K_\max$ achieved by real rays
($\gamma_3$ and $\gamma_4$ in Fig.~\ref{fig:ellipseevan})
escaping horizontally along directions of least intensity. It can
be shown that these latter rays are launched from points
\[
u=\pm\frac{\pi}{2}+\rmi U_2,\quad\mbox{where}\quad
\cosh U_2=\sqrt{\frac{n^2M}{a^2-n^2c^2}}
\]
and have an imaginary action $\Im\,S=K_\max=a\sinh U_2$.

Following this change of coordinates, the Herring integral
is left in the form
\begin{eqnarray}
\label{eq:unifint}
\kappa=\frac{1}{2\pi}\int_{0}^{2\pi} \tilde{A}(\chi)
\rme^{-k_r(\bar{K}+\Delta K \cos2\chi)}\textrm{d}\chi.
\end{eqnarray}
We next write
\begin{eqnarray}\label{giveA}
\tilde{A}(\chi)=A_0+A_1\cos2\chi+H(\chi)
\frac{\partial}{\partial\chi}\Delta K\cos2\chi,
\end{eqnarray}
where $A_0$ and $A_1$ are constants, chosen so that
\begin{eqnarray*}
H(\chi)=\frac{A_0+A_1-\tilde{A}(\chi)}{2\Delta K\sin2\chi}
\end{eqnarray*}
is smooth, which (exploiting the symmetry of the
problem with respect to $u\to u+\pi$) is true provided
\begin{eqnarray}\label{A1}
\tilde{A}(0)=A_0+A_1
\end{eqnarray}
and
\begin{eqnarray}\label{A2}
\tilde{A}\left(\frac{\pi}{2}\right)
=A_0-A_1.
\end{eqnarray}
Substitution of (\ref{giveA}) in (\ref{eq:unifint}), allows us,
following integration by parts, to neglect at leading order
in $1/k_r$ the contribution of the last term in (\ref{giveA}). The
result is
\begin{equation}\label{bessels}
\kappa
\approx(A_0 I_0(k_r\Delta K)+A_1I_1(k_r\Delta K))\rme^{-k_r\bar{K}},
\end{equation}
where $I_0(z)$ and $I_1(z)$ denote modified Bessel functions of the first kind.

In practice, the amplitudes $A_0$ and $A_1$ are conveniently
evaluated by comparing the large-deformation asymptotics
of (\ref{bessels}) with the amplitude terms of
Wilkinson's formula. A straightforward comparison yields
\[
(A_0+A_1)\sqrt{\frac{\pi}{k_r\Delta K}}
=\frac{1}{(2\pi )^{3/2}k_r^{1/2}} \frac{|t(u_1)|^2\J}{\sqrt{ |\{M,M^*\}_{u_1}}|}.
\]
Although the physical problem insists that $k_r$ is positive,
we might also formally compare asymptotic results for negative
$k_r$ and this yields
\[
(A_0-A_1)\sqrt{\frac{\pi}{k_r\Delta K}}
=\frac{1}{(2\pi )^{3/2}k_r^{1/2}}   \frac{|t(u_3)|^2\J}{\sqrt{ |\{M,M^*\}_{u_3}}|}.
\]
The right-hand side here is determined by the geometry of the ray family
around real rays escaping along the directions of \textit{least}
emitted intensity. Here we need the dark analogue of (\ref{brightPB}):
\begin{equation}
\rmi\{M,M^*\}_{u_3,u_4}=-\frac{8(n^2-1)c^2a}{n^2b^2}
M\sqrt{\frac{n^2M-a^2+n^2c^2}{a^2-n^2c^2}}.
\end{equation}
$A_0$ and $A_1$ are then known and the evaluation of (\ref{bessels})
is complete.

\begin{figure}
  \centering
    \includegraphics[width=0.75\textwidth]{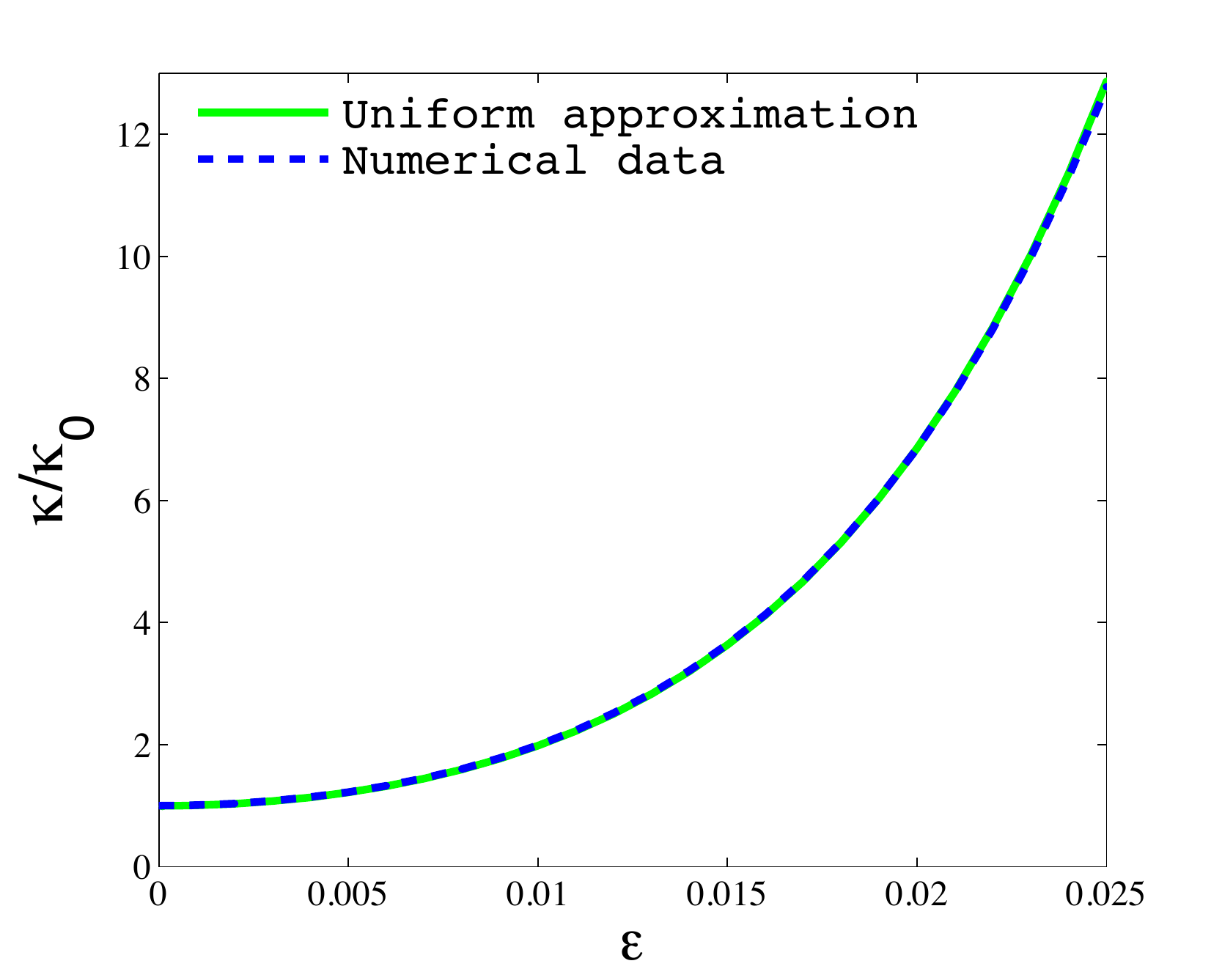}
    \caption{A comparison is given between calculations of $\kappa$
using  the uniform result (\ref{bessels}) and a numerical evaluation for the
same mode and parameter ranges used in Fig.~\ref{fig:kappawilk}.}
    \label{fig:kappabes}
\end{figure}

This result is compared with numerical data in Fig.~\ref{fig:kappabes}
and is seen to give a very good description of the $Q$-factor degradation
from the perturbative regime of (\ref{genresult}) to the
larger-deformation regime of (\ref{eq:kappawilk}). In practice, the
deformations treatable by this result are limited by our ability (or
otherwise) to find phases $\sigma$ --- which have been absorbed
in the transmission factor $|t|^2$ in this paper --- which account
for the dielectric boundary conditions in this problem. Such phases
have been seen in \cite{creaghellipse10} to suffer from natural
boundaries which approach the real axis as eccentricity increases
and invalidate the WKB ansatz above a critical deformation in the region
of the complex boundary from which escaping rays are launched. It
should be emphasised that the eccentricities at which this happens
are formally of $O(1)$ and much greater than the eccentricities
at which the perturbative result fails.

\section{Conclusions}
\label{sec:conc}

We have found that, in a regime of short wavelength, very small
deformations of optical cavities suffice to alter significantly
the quality factor of resonant modes and have given quantitative
estimates of this effect. We have shown that the quality factor
may be halved or worse by deformations as small as one
part in a thousand, for example, in explicit calculations
where the cavity is some  tens of wavelengths
in diameter.

Although complex WKB methods provide
a natural starting point for such calculations, their implementation is
problematic because the required ray data may be difficult to find, or
may not even exist. Instead, we have provided a general analysis
based on approximation of the underlying ray families using
perturbation techniques. Although restricted to relatively weak
deformations, because the perturbations are applied to the rays rather
then directly to the wave solution itself, this approach can
nevertheless successfully describe deformations where the wave
and quality factor itself are altered nonperturbatively.

For moderately large deformations, the results of this analysis are
naturally interpreted in terms of the geometry of families of
escaping rays, using Wilkinson's formula (\ref{eq:kappawilk}).
Deformation effects the important qualitative change of making the
associated family of escaping rays slightly complex (whereas
far enough away from the cavity the corresponding orbit families
are real in the undeformed case).
Nevertheless a discrete subset of the escaping rays remain real
and these in particular determine the directions along which the
waves of greatest intensity propagate. Wilkinson's formula
expresses the quality factor degradation in terms of the geometry
of the ray family around the real subset guiding waves of highest
intensity. It fails in the limit of small
deformation where a denominator, which measures the rate at which
rays become complex away from the real subset, vanishes. A uniform
result has been also derived, however, which interpolates between the
primitive Wilkinson formula appropriate to larger deformations and
the perturbative analysis derived earlier for weak deformations.
This was shown to
give an excellent description of the quality factor in the special case
of elliptical deformations, where the ray families can be calculated
analytically even for significantly eccentric cavities.

\ack{The authors are grateful to Martin Sieber for useful discussions.
This work was supported by EPSRC under grant number EP/F036574/1.}

\section*{References}
\bibliographystyle{ieeetr}
\bibliography{mwbibliography}
\end{document}